\documentclass[aps,prb,reprint,superscriptaddress]{revtex4-1}
\usepackage[applemac]{inputenc}	
\usepackage[T1]{fontenc}
\usepackage[cyr]{aeguill}
\usepackage[pdftex]{graphicx}
\usepackage{wrapfig}
\usepackage[english,french]{babel}
\usepackage{amsmath}
\usepackage{amssymb,amsfonts,textcomp}
\usepackage{color}
\usepackage{array}
\usepackage{hhline}
\usepackage{hyperref}
\usepackage{fancyhdr}
\usepackage[pdftex]{graphicx}
\definecolor{violet}{rgb}{0.5,0,0.5}
\definecolor{violet2}{rgb}{0.61,0,0.87}
\definecolor{vert}{rgb}{0,0.65,0}
\definecolor{vert2}{rgb}{0,0.6,0}
\definecolor{marron}{rgb}{0.79,0.52,0}
\definecolor{orange}{rgb}{0.87,0.47,0}
\definecolor{lightblue}{rgb}{0,0.8,0.8}
\definecolor{jaune}{rgb}{0.948,0.91,0.17}

\begin{document}
\thispagestyle{empty}
\title{Time- and Space-Modulated Raman Signals in Graphene-based Optical Cavities}
\author{Antoine Reserbat-Plantey}\altaffiliation{Present address : ICFO-Institut de Ciencies Fotoniques, Mediterranean Technology Park, 08860 Castelldefels (Barcelona), Spain}
\affiliation{Institut N\'eel, CNRS-UJF-INP, 38042 Grenoble Cedex 09, France.}
\author{Svetlana Klyatskaya}\affiliation{Institute of Nanotechnology, Karlsruhe Institute of Technology (KIT), Germany.}
\author{Valérie Reita}\affiliation{Institut N\'eel, CNRS-UJF-INP, 38042 Grenoble Cedex 09, France.}
\author{La\"etitia Marty}\affiliation{Institut N\'eel, CNRS-UJF-INP, 38042 Grenoble Cedex 09, France.}
\author{Olivier Arcizet}\affiliation{Institut N\'eel, CNRS-UJF-INP, 38042 Grenoble Cedex 09, France.}
\author{Mario Ruben}\affiliation{Institute of Nanotechnology, Karlsruhe Institute of Technology (KIT), Germany.}\affiliation{IPCMS-CNRS-Université de Strasbourg, 67034 Strasbourg, France.}
\author{Nedjma Bendiab}\affiliation{Institut N\'eel, CNRS-UJF-INP, 38042 Grenoble Cedex 09, France.}
\author{Vincent Bouchiat}\affiliation{Institut N\'eel, CNRS-UJF-INP, 38042 Grenoble Cedex 09, France.}

\begin{abstract}
 We present fabrication and optical characterization of micro-cavities made of multilayer graphene (MLG) cantilevers clamped by metallic electrodes and suspended over Si/SiO$_2$ substrate.
Graphene cantilevers act as a semi-transparent mirrors closing an air-wedge optical cavity.
This simple geometry implements a standing-wave optical resonator along with a mechanical one. 
Equal thickness interference fringes are observed in both Raman and Rayleigh backscattered signals with interfringe given by their specific wavelength. 
 Chromatic dispersion within the cavity makes possible spatial modulation of graphene Raman lines and selective rejection of the silicon background signals. 
Electrostatic actuation of the multilayer graphene cantilever by gate voltage tunes the cavity length and induces space and time modulation of backscattered light including Raman lines. 
We demonstrate the potential of those systems for high sensitivity Raman measurements of generic molecular species grafted on multilayer graphene surface. 
The Raman signal of the molecular layer can be modulated both in time and in space in a similar fashion and show enhancement with respect to a collapsed membrane.

\end{abstract}
\date{\today}
\maketitle
\clearpage
\textbf{}

\section{Optical Properties of Graphitic systems }

\subsection{Introduction: interaction of light with graphene}

Graphitic carbon nanostructures are generally known to produce very efficient light-absorbing media\cite{takadoum2010black}. 
Indeed the "blackest" materials ever manufactured are based on carbon nanotube forests. 
Interestingly the exact same carbon phases when made very thin and flat (such as in carbon nanotube thin films or stacked graphene layers), provide extremely promising candidates\cite{kim2009large} for the implementation of flexible and transparent electrodes. This apparent paradox can be easily explained by looking at the variety of structures and shapes into which nanostructures based on sp$^2$ hybridized carbon can be assembled down to the atomic level. 
Among all these allotropes, graphene is known for providing a flat 2D material with outstanding optical, electrical and mechanical properties\cite{Cooper2012}. 
Optics in graphene had an historical role as it was strongly involved in the first isolation of a monolayer \cite{Novoselov2004,Novoselov2005,Zhang2005}.  
Indeed a single layer of graphene can be easily seen under an optical microscope as the reflected light has  a sufficient phase shift \cite{Blake2007} to be detected by the bare eye. 
Optical methods based on reflectometry have been identified as critical to efficiently sort graphene multilayers (MLG) obtained by exfoliation techniques as it readily allows the quantitative identification of the number of stacked layers~\cite{Roddaro2007}. 
Moreover owing to its peculiar quantum electronic properties, graphene acts as a semi-transparent membrane exhibiting universal and quantized optical transmission\cite{Nair2008}. 
This quantization fails for multilayers graphene thicker than a few layers, however the membrane keeps a significant transparency up to hundred of layers\cite{Skulason:2010}.  
 
\subsection{Enhancing graphene coupling to photons by integration within an optical cavity}

\begin{figure}[htbp]
\begin{center}
		\includegraphics[width=0.35\textwidth]{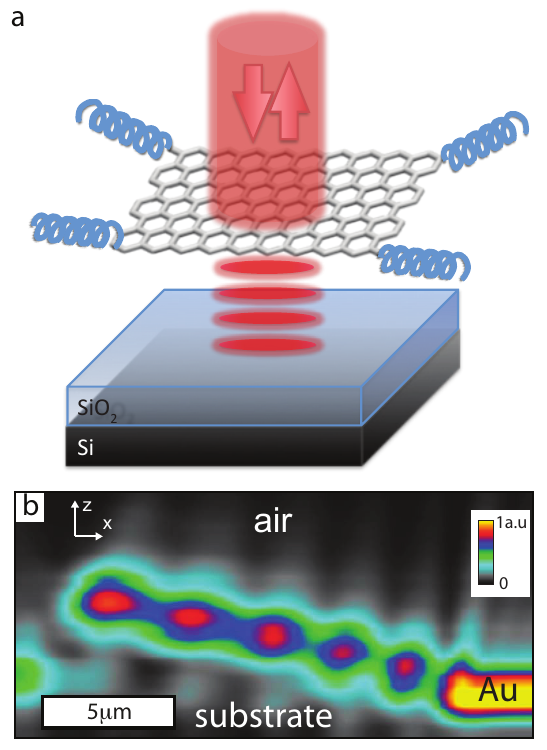}
	     \caption{\textbf{Graphene based optomechanical systems.} \textbf{a} : a mechanically flexible and semi-transparent graphene-based membrane is freestanding over a silicon back-mirror, constituting an optical cavity. Elastically  and inelastically back-scattered light is measured and gives indications on the both the cavity thickness and the photonics interaction with bottom and top mirrors. \textbf{b}: Spatially-resolved reflectance map acquired with a conical microscope in the cross sectional (X,Z) plane at 532 nm for typical sample. This experiment has been performed using scanning confocal microscopy at different focus showing the reflected intensity modulation by optical interference within the cavity.}\label{fig2}
\end{center}
\end{figure}

In all these previous studies, optical interferences is at the origin ofthe visibility of graphene and more generally gives means to further optimize  the interaction with incoming photons.  
Light-graphene coupling can be indeed further enhanced by integrating the graphene within an optical cavity. 
This has been recently achieved on photonic crystals~\cite{GuT:2012p4107} or in a Fabry-Perot cavity~\cite{furchi2012microcavity,Engel:2012p4108}. 
 it was shown that the sensitivity of graphene-based photon detectors\cite{doi:10.1021/nl3016329,Furchi2012,Engel2012} can be maximized in a cavity in which graphene is placed at resonant conditions,. 
For example thanks to the cavity effect, the absorbance of graphene in the THz range can exhibit a wavelength absorbance contrast reaching up to 70\% modulation depth\cite{caviteTHz}.  Recently, a suspended and movable graphene membrane has been integrated into a low finesse optical cavity\cite{Barton2012}. 
This demonstrates how light can be coupled to the mechanical motion, and lead to active cooling of the graphene membrane, showing the interesting prospects of using graphene for further optomechanical studies (cf. figure \ref{fig2}).
Similar cavity effects also apply to Raman scattered light.
Raman spectroscopy, a powerful probe for graphene as it allows the local mapping of numerous physical parameters such as doping, structural defects, phonon temperature and mechanical stress, can be strongly affected by its coupling to a cavity. 
On this basis, interference enhancement of Raman scattering (IERS) has been studied long ago\cite{Connell1980} and more recently applied to graphene\cite{Jung2010,Yoon2009} by tuning the SiO$_2$ thickness of samples made of supported or suspended graphene flakes. 
Similarly to what was achieved for elastically scattered light, IERS relies on the constructive interference of Raman scattered photons \cite{Connell1980,Fainstein1995}.
Multiple wave interferences in graphene over silica have been identified as a way to enhance the Raman signals either coming from the graphene vibrations modes\cite{Yoon2009} or from adsorbates such as halogen species\cite{Jung2010}. 
By varying the composition\cite{Castriota2010608} and thickness \cite{Yoon2009} of the dielectric underlayer on which the graphene is deposited, modulation of the Raman lines has been shown.

In the present work, we show that, by building an optical micro-cavity made of self-supported multi-layer graphene cantilever freestanding over silicon, interference phenomena occur not only in back-reflected light but in Raman signals, either coming from graphene, from silicon substrate or from any foreign adsorbed element. By electrostatic actuation of the graphene membrane, a time modulation synchronized with the application of a AC signal on the metallic electrodes can be achieved, offering a versatile platform for Raman studies. 
 
\section{Fabrication and Implementation of Multilayered graphene electrically actuated optical cavities}

Our system takes advantage of the outstanding  optical, mechanical and electrical properties of graphene. Multilayers of typically 10 to 100 layers implement suspended semi-transparent mirror cantilevers, stiff enough to remain self-supported over more than 10 microns and electrically conducting to be actuated by an external electrostatic potential.  The simultaneous high rigidity of this material and its semi-transparency allow creating an overhanging cantilever which is at the same time, light, rigid and semi-transparent (typically $\sim$ 25 \% transparency, $\sim$ 30 \% reflectance, $\sim$ 45 \% absorbance for 100 monolayers of graphene\cite{Skulason:2010}). 
The batch integration of this system can be made using exfoliation of graphite on oxidized silicon followed by lithography/deposition of electrodes and under etching (see methods for details). 

\begin{figure}[htbp]
\begin{center}
		\includegraphics[width=0.48\textwidth]{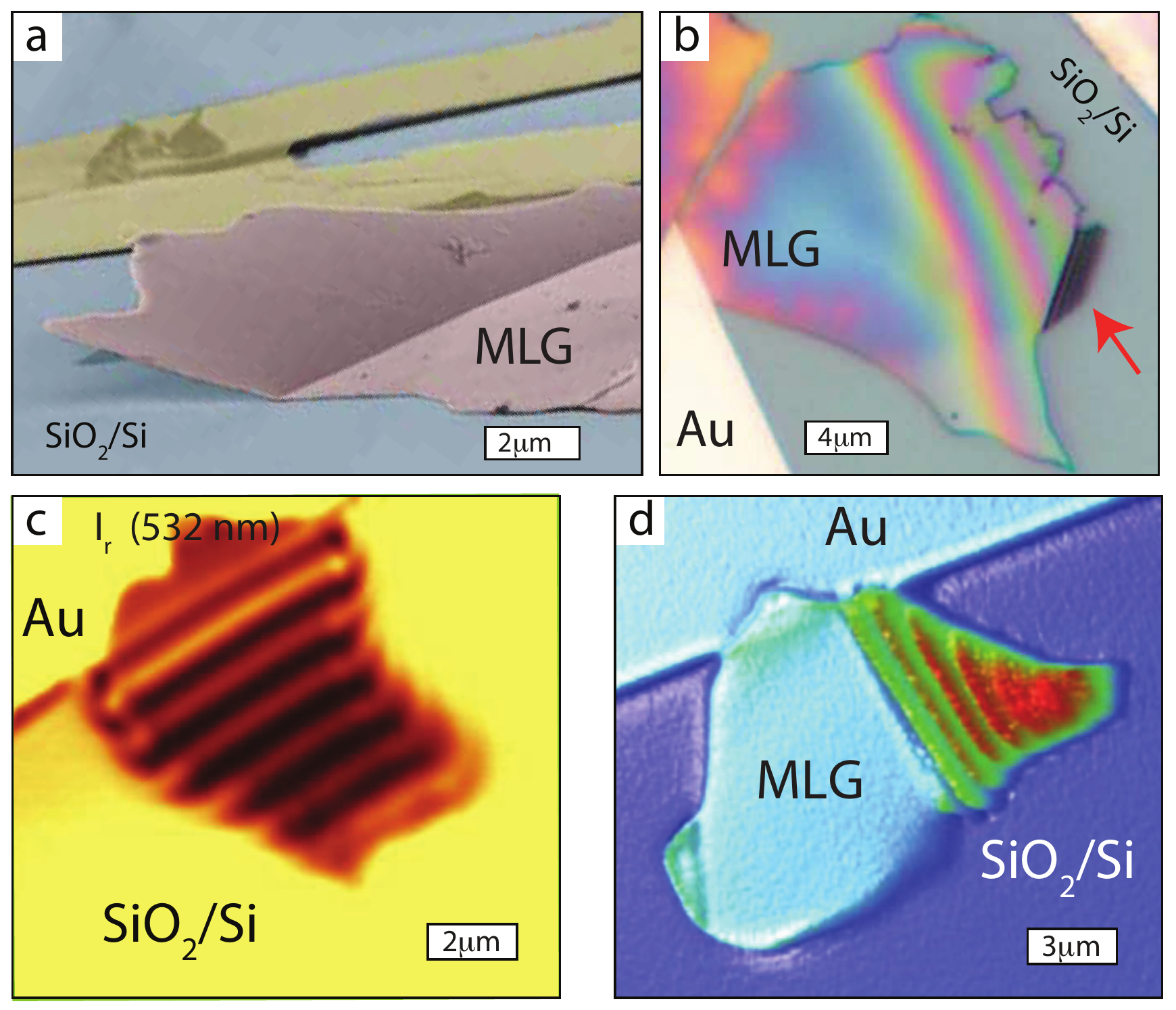}
	     \caption{\textbf{Multilayer graphene optical cavities.} \textbf{a}:  Scanning electron micrograph of multilayer graphene flake overhanging the silicon oxide defining an air wedge optical cavity. \textbf{b}: Top view optical micrograph under white light illumination showing iridescence. \textbf{c}: Reflectance confocal (X,Y) map obtained by scanning a 532 nm laser probe, showing the equal thickness interference fringes. \textbf{d}: Phase micrography recorded with an optical profilometer revealing the interference pattern along the wedged membrane.}\label{fig1}
\end{center}
\end{figure}
This optical micro-cavity defines an air wedge structure as shown on figure \ref{fig1}.
Unlike conventional\cite{Ling2010} graphene-based optical cavities with fixed geometries, the optical length of this cavity increases linearly along the cantilever, which allows the observation of multiple interference fringes (cf. figure \ref{fig2}).
These equal-thickness interferences fringes, so-called Fizeau fringes, will be first described for reflected light and Raman scattering. 
Furthermore, we demonstrate the possibility to tune this optical cavity with an voltage applied on graphene, allowing us to control the wedge angle and thus the interference pattern. The resulting shift in the interference pattern can be used a an intensity modulation for a fixed laser probe, which in turn can be used to modulate the intensity of Raman signals.  
Taking advantage of the variable interference enhancement of Raman scattering combined with the ability of graphene to be an excellent substrate for Raman scattering, we show the possibility to use this cavity as a tunable platform for molecular detection.
Recently, our team has shown that similar suspended graphitic membranes can be used to simultaneously detect motion and stress\cite{Reserbat-Plantey2012} within the actuated cantilever using the stress-dependence of the Raman response.
Perspectives for the use of these novel families of system for building hybrid platforms for Raman studies will be presented. 

\section{Optical Fizeau interferences in the micro-cavity}
\subsection{Interferences in reflection}
\begin{figure}[htbp]
\begin{center}
		\includegraphics[width=0.48\textwidth]{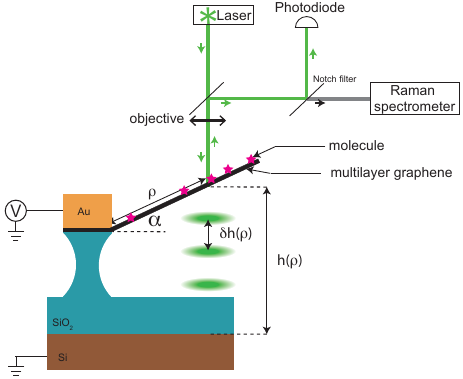}
	     \caption{\textbf{Schematics of the measurement setup.} A multilayer graphene cantilever clamped between silica pillar and a gold electrode is addressed by a gate voltage that controls its wedge angle $\alpha$ which is probed by reflectometry (signal measured with high bandwidth photodiode) and Raman spectrometry (filtered inelastic scattering signal measured with spectrometer coupled to a CCD device). The height between two successive fringes in reflection is labeled $\delta h=\lambda/(2n_0)$. Raman-active molecular species (red stars) adsorbed on the mirror can be detected as well.
}\label{fig5}
\end{center}
\end{figure} 
Optical interference fringes are observed along this air wedge device (cf. figure \ref{fig1}b-d and \ref{fig2}b).
As the local cavity thickness $h(\rho)$ (which corresponds to the vertical gap between the MLG membrane and the reflecting substrate) is increased by $\delta h = \lambda/(2n_0)$, where $\lambda$ and $n_0$ are respectively the wavelength and the optical index of air or vacuum (cf. figure \ref{fig5}), a new interference maximum appears within the cavity. 
In fig \ref{fig1}b-c, iridescence is observed under white light illumination and the interference pattern is measured at 532 nm. 
Wedge geometry implies that all points at equal distance from the hinge have equal light wave dephasing, forming a set of parallel Fizeau fringes (or equal thickness fringes).
Note that the interference order $2h/\lambda$ is close to one as the thickness of the wedge are a mere  $\lambda$/2, especially near the hinge position. 
In addition, for small wedge angles ($\alpha \sim$10$^{\circ}$), the in-plane distance between two adjacent fringes would be large enough to spatially resolve them by confocal microscopy imaging.
By confocal imaging, AFM and profilometry measurements, we check that the geometry of the MLG cantilever is flat. 
This ensures that the gap between the MLG cantilever and the silicon substrate increases linearly from the hinge to the end of the cantilever.
One can simply consider light beam enters the cavity and get trapped for a certain number of reflexions on both sides before escaping it. 
For high finesse cavity, this trapping becomes efficient and light field can be locally strongly enhanced inside the cavity.
In our case, the wedged cavities (finesse $\sim$ 10) do not exhibit the high finesse characteristic of Fabry-Perot fringes i) because of the non-parallel wedge geometry which causes wave detrapping at higher interference order\cite{Rogers1982}, and ii) because light absorption limit is not negligible in both substrate and MLG membrane.
The first argument, previously discussed by Brossel\cite{Brossel1947}, can be interpreted in terms of geometrical optics.
Reflexions of the multiple reflected beam in the cavity is progressively shifted along the membrane, eventually escaping the cavity. 
Therefore, after a number $p$ of reflexions given by the Brossel criterion $p\le \left( 3\lambda/(8 h \alpha^3) \right )^{1/3}$ , the transmitted beam cannot be collected by the microscope objective.
Such effect implies i) a low intensity of the interference fringe, ii) an increase of its width along the membrane, iii) a shift of the fringe along the wedge structure and iv) an asymmetry in the fringe profile.
The second reason why those cavities do not exhibit very high finesse is the strong absorption of light which limits the reflected intensity $I_{reflected}$.
Interestingly, the derivation of the reflected light in such asymmetric cavity becomes highly non-trivial problem when absorption effect has to be considered as highlighted by Holden\cite{holden1949} in 1949.
According to these two effects, one may derive the reflected light intensity : 
\begin{equation}
I_{reflected}[\rho] \sim I^0 . \left [ \frac{\mathcal{F}}{2\mathcal{A}} (1+\cos \phi_{laser} [\rho]) +\Gamma\right ],
\label{eq:gain}
\end{equation}
where $I^0 $, $\rho$, $\mathcal{F}$, $\mathcal{A}$, $\phi_{laser} [\rho]$ and $\Gamma$ are respectively the incident laser light intensity, the laser position along the membrane, the cavity finesse, the absorption coefficient, the phase difference between two reflected beams and a constant which slightly depends on the reflection coefficient\cite{holden1949}.
One can express the phase difference $\phi_{laser} [\rho]=\frac{2\pi}{i_{laser}}\rho$ as a function of the interfringe $i_{laser}$ which is the distance separating two successive Fizeau fringes along the MLG membrane.

Raman effect is an inelastic scattering process of light and gives information on the atomic displacements inside a material. 
Analysis of the probed optical phonons leads to a rich variety of intrinsic properties (thermal, mechanical, electronic, vibrational, structural) \cite{Cooper2012,Ferrari2006,Lopes2010,Reserbat-Plantey2012,Yoon2009}. 
Moreover, since Raman signal is related to the microscopic structure of matter, it is a fingerprint of a lot of chemicals or molecule. 
Raman fingerprint of graphitic nano-system is made of two major G and 2D bands\cite{Ferrari2006}.

\begin{figure}[htbp]
\begin{center}
		\includegraphics[width=0.35\textwidth]{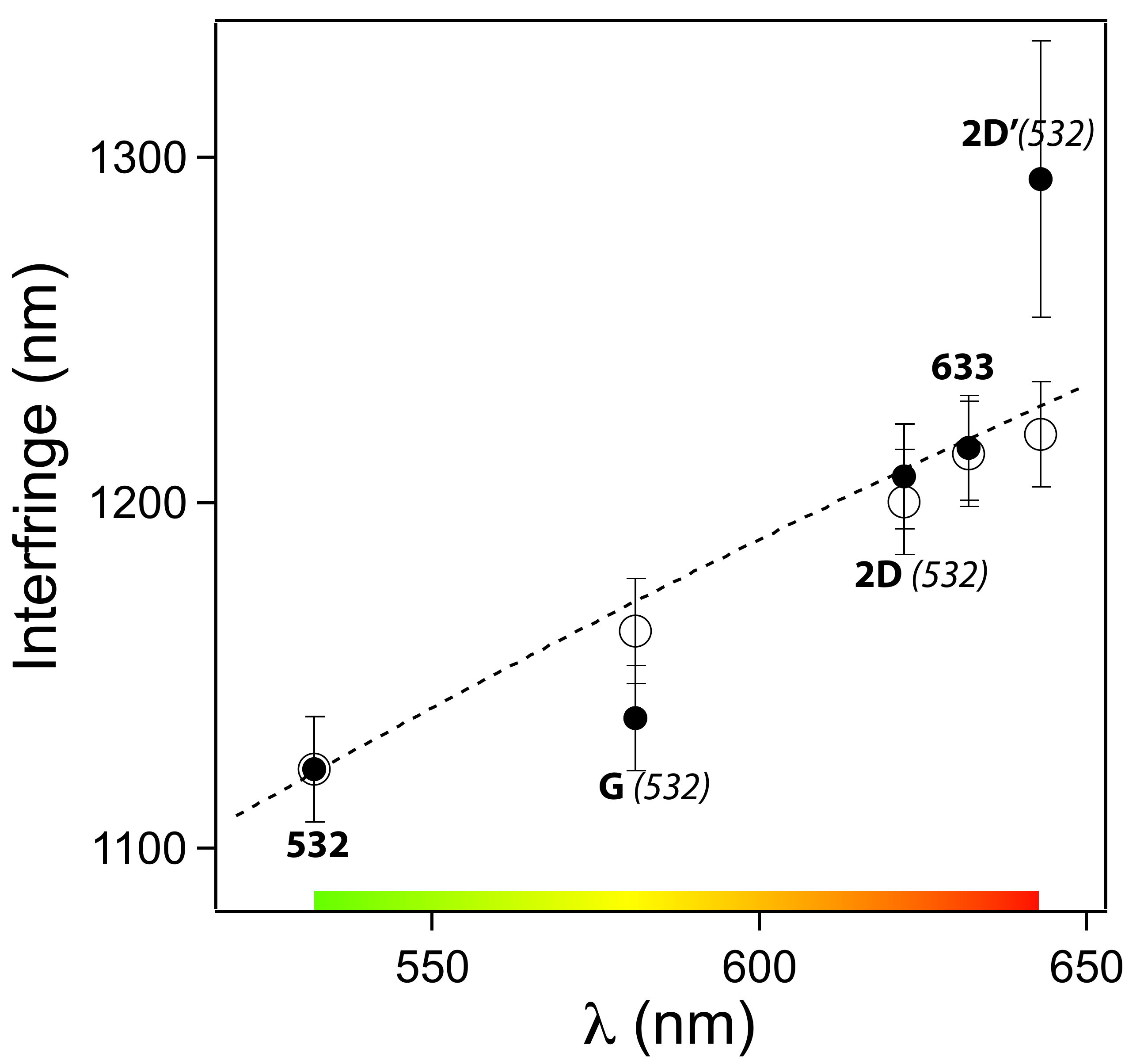}
	     \caption{\textbf{Spatial modulation of intensity of both reflected and Raman lines.}  Measurement of the inter-fringe width (averaged over the sample) as a function of the effective photon wavelength. Experimental data ($\bullet$) are in agreement with basic interferometric simulation ($\circ$) performed with Nanocalc, and follow a quasi-linear dispersion (dashed line) as suggested by the equation \ref{eq:interfringeRa}. Points at 532 nm and 633 nm correspond to the light reflection for the two excitations, while MLG Raman bands are labeled as G, 2D and 2D' with the associated pump laser into brackets.}\label{fig4}
\end{center}
\end{figure}

\begin{figure}[htbp]
\begin{center}
		\includegraphics[width=0.40\textwidth]{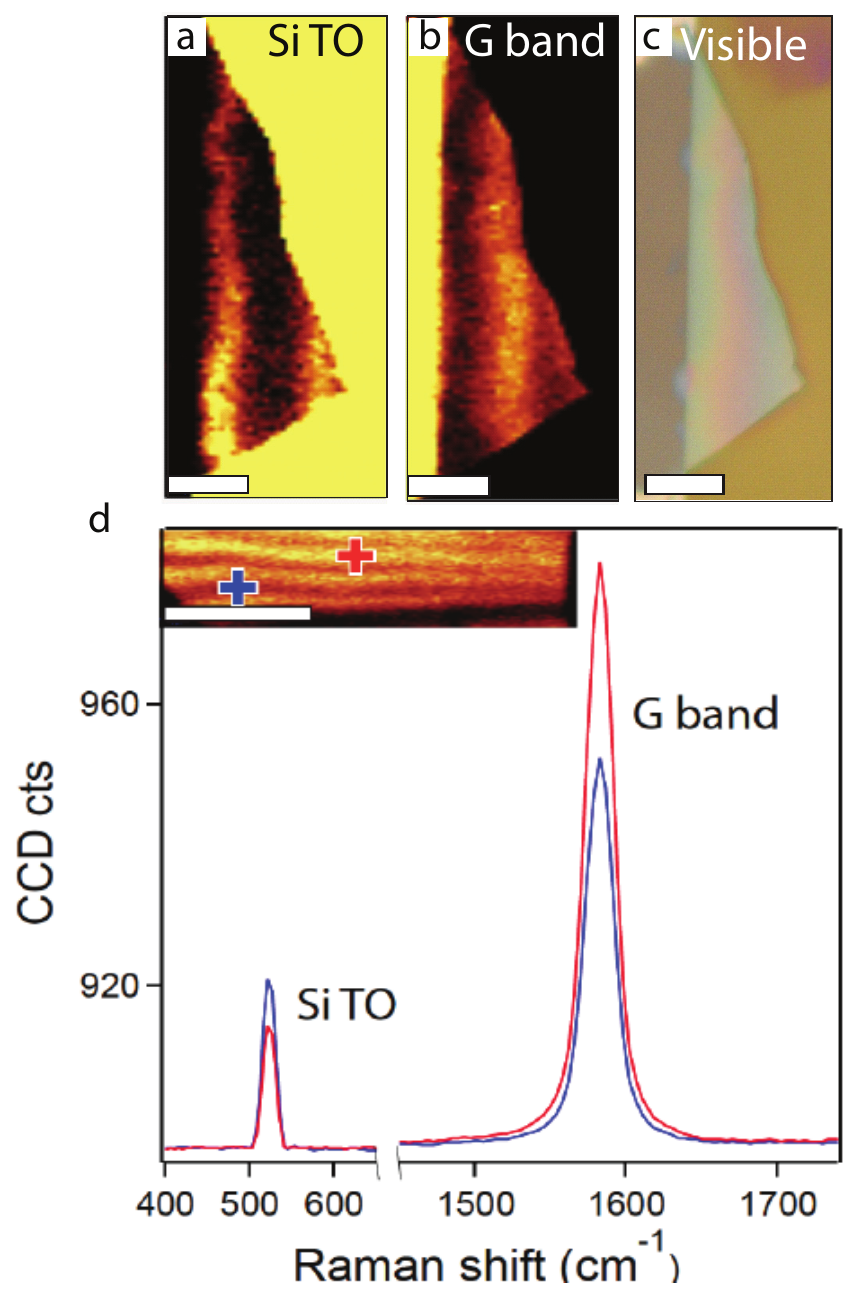}
	     \caption{\textbf{Multi-Layer-Graphene wedge structure overhanging silicon oxide.} \textbf{a-c}: Raman intensity maps of the silicon representative mode (Si-TO) and the G band are presented as well as the optical image of the corresponding device. Each mode presents its own interference pattern which is clearly shifted due to chromatic dispersion within the cavity. \textbf{d}: Raman spectra showing Si-TO and G bands taken at two different locations along the suspended MLG cantilever. Red curve is taken at a bright fringe of the G band intensity pattern, while blue is taken at a dark fringe position. Inset : G band intensity profile with positions were the two spectra are taken. Scale bars represent 10 $\mu$m. }\label{fig3}
\end{center}
\end{figure}

Interference patterns observed along MLG cavities from the reflection of the pump laser are also observed from the Raman scattered light and follow a similar behavior (see figures \ref{fig3} and \ref{fig4}). 
Thus, one can see a situation where light sources are now localized on one side of the cavity (either the Si or MLG side).
The light sources create Raman photons which can enter the cavity and produce interference pattern, or get collected directly by the objective. 
Interestingly, not all the collected Raman photons contribute to interference effect since the collection angle is very large (NA = 0.95), and thus the collected Raman light has to contribution : i) interference enhanced Raman (IERS) and ii) standard Raman.
Equation \ref{eq:gain} is then modified, and we define $I_{Raman}$ as the Raman intensity (i.e. the G band) measured along the membrane :
\begin{equation}
I_{Raman}[\rho] = I_{Raman}^0[\rho]  \left \lbrace \frac{\Omega_1}{4\pi} + \frac{\Omega_2}{4\pi}\left[ \frac{\mathcal{F}}{2} \left( 1-\cos \phi_{Ra}^0[\rho] \right) \right] \right \rbrace
\label{eq:Raman_interf}
\end{equation}
Where $I_{Raman}^0[\rho]$ is the total Raman scattered intensity at position $\rho$ and for a given laser power, without taking into account interference effects.
$\Omega_1 = 2\pi \left (1-\cos(\sin^{-1}NA)\right )$ is the solid collection angle for photons without interference contribution, and $\Omega_2$ is the solid collection angle for photons interfering in the cavity.
$\phi_{Ra}^0[\rho] = \frac{2\pi}{i_{Ra}^0}\rho$ is the phase difference between interfering beams and $i_{Ra}^0$ is the theoretical interfringe for a given Raman mode in case of homogeneous laser power.
For isotropic scattering, the fist term in equation \ref{eq:Raman_interf} is independent of the position along the membrane and only depends on the numerical aperture. 
In our case, $\Omega_1/4\pi \sim 34 \%$.
The IERS signal (collection solid angle $\Omega_2 = 2\pi (1-\cos\theta_2)$) obeying the Brossel criterion can be understood as the maximal interference order allowed for a light beam transmitted out of the cavity at a position laterally shifted by a distance smaller than the laser spot size. 
Limit angle $\theta_2 = \tan^{-1} \left [ \lambda_{ph}/(4h[\rho]) \right ]$ is derived for a two-waves interference. 
Thus, $\Omega_2$ depends on the air wedge thickness $h[\rho]$ : close to the hinge $\Omega_2$ can correspond to $1/3$ of $\Omega_1$ ($h[0] \sim$ 180 nm, $\alpha \sim$ 12 $^{\circ}$.).

So far, we have introduced the intensity modulation for laser light  (cf. eq. \ref{eq:gain}) and for Raman scattered light (cf. eq. \ref{eq:Raman_interf}).
In equation \ref{eq:Raman_interf}, we have introduced the term $I_{Raman}^0[\rho]$, which depends on the laser position along the membrane since we take into account a realistic case where Raman scattered light intensity is affected by the modulation of the laser excitation intensity.
This leads to : 
\begin{widetext}
\begin{equation}
I_{Raman}[\rho] = \frac{\mathcal{F} I^0_{Raman}}{8\pi}\left \lbrace \Omega_1 \left[ 1-\cos\left ( \frac{2\pi}{i_{laser}}\rho\right )\right]+\frac{\Omega_2 \mathcal{F}}{2} \left [ 1-\cos\left ( \frac{2\pi}{i_{laser}}\rho\right )\right  ]\left [ 1-\cos\left ( \frac{2\pi}{i_{Raman}^0}\rho\right )\right  ]\right \rbrace
\label{eq:Iramandetail}
\end{equation}
\end{widetext}
It is worth noting that equation \ref{eq:Iramandetail} shows the non-trivial dependence of the observed Raman Fizeau interfringe $i_{Raman}$ since it is now clear that there is beating effect due to interferometric modulation of laser excitation.
Three interfringes expressions are present : the measured Raman interfringe $i_{Raman}$, the absolute Raman interfringe (supposing constant laser power along the membrane) $i_{Raman}^0=\lambda_{Raman}/(2\alpha n_0)$, and the excitation laser interfringe $i_{laser}\lambda_{laser}/(2\alpha n_0)$.
From the equation \ref{eq:Iramandetail}, two regimes appears : i) $\Omega_1 \gg \frac{\Omega_2\mathcal{F}}{2}$ (Raman interference effect negligible) leading to $i_{Raman} \sim i_{laser}$, and ii) $\Omega_1 \ll \frac{\Omega_2\mathcal{F}}{2}$(Raman interference effect dominant) leading to one physical solution :
\begin{equation}
\begin{split}
\left ( i_{Raman}\right )^{-1} &= \frac{1}{2}\left [ \left ( i_{Raman}^0\right )^{-1} + \left ( i_{laser}\right )^{-1}\right ]\\
i_{Raman} &= \frac{1}{\alpha n_0} \ \frac{\lambda_{Raman}\lambda_{laser}}{\lambda_{Raman}+\lambda_{laser}}.
\label{eq:interfringeRa}
\end{split}
\end{equation}
This expression, plotted in figure \ref{fig4}, fits the experimental data and the simulated points using an ellipsometry simulation software. 
In figure \ref{fig4}, we clearly see that $i_{Raman}\ne i_{laser}$ and depends on the Raman scattered light wavelength. 

In figure \ref{fig3}a-c, interference fringes are measured \textit{via} spatially resolved Raman spectroscopy for silicon (TO) and MLG (G, 2D and 2D') Raman lines. 
By taking advantage of this periodic shift with respect to $\lambda$ (typically $\rm{\sim100 \ nm}$), the $\rm{I_G/I_{Si}}$ ratio can be tuned by moving the probe position (see Fig. \ref{fig3}d), in order to optimize the spatial modulation and background contrast.
Such a modulation (by moving the laser) allows to tune the optical cavity to maximize a Raman peak relative to another which could be of great use in high resolution Raman spectroscopy.
This effect is even more pronounced if the finesse of the optical cavity is high (as suggested by equation \ref{eq:Iramandetail}) and offers the possibility of effective rejection of unwanted Raman signals (substrate, residues, etc.).
Another manner to modulate this Raman signal is to realize an electrostatic modulation.

\section{Electrostatic modulation of the optical micro-cavity}

The optical cavity thickness can be adjusted by varying the angle $\alpha$ through electrostatic actuation of the cantilever (see figure \ref{fig5}), thus producing a shift in the interference pattern. 
This modulation of the cavity thickness $h(\rho)$ is achieved by applying a DC or AC voltage applied to the clamp electrode while the substrate is grounded. 
Thus $h(\rho,t) = h(\rho)_0 + \delta h(\rho,t)$ where $h(\rho)_0$ is the local cavity thickness or length at $V=0$.
This results in an attractive electrostatic force: 
\begin{equation}
\vec{F} = - \dfrac{1}{2}\dfrac{\partial C_{tot}(h)}{\partial h} V^2\vec{u_{\theta}} = - B[h(\rho)_0]V^2\vec{u_{\theta}}.
\label{eq:forceElec}
\end{equation}
This force (eq. \ref{eq:forceElec}) produces an angular deviation with respect to the equilibrium position. 
For driving frequencies $\Omega << \Omega_m$ where $\Omega_m$ is the mechanical resonant frequency (MHz range\cite{Reserbat-Plantey2012}), the motion of the cantilever of mass M is determined by linear response theory and $h(\rho,t)$ is then defined as the product of the mechanical susceptibility at low frequency $\chi_{mac} = M/\Omega_m^2$ and the applied force $F(t)$ : 
\begin{equation}
h(\rho,t)  = \chi_{mec} F(t) = \chi_{mec} B[h(\rho)_0] \ V(t)^2.
\label{LW}
\end{equation}
It is worth noting when laser spot is located 10 $\mu$m away from the hinge, the deviation $\delta h(\rho,t)$ does not change significantly ($< 1\%$) the value of $B[h(\rho)_0]$ which could be approximated as a constant. 
 
\begin{figure}[htbp]
\begin{center}
		\includegraphics[width=0.48\textwidth]{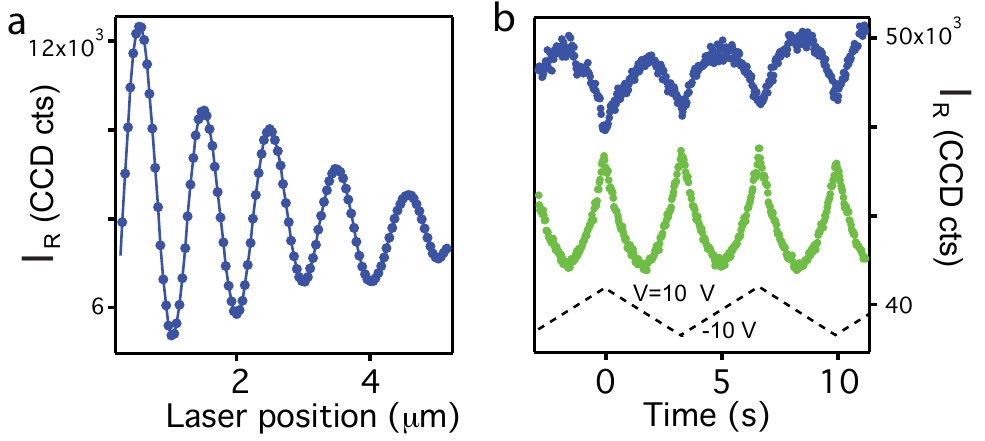}
	     \caption{\textbf{Spatial and time modulation of the backreflected signals.} \textbf{a} : Reflected light intensity ($\lambda_{laser}$ = 532 nm) along the MLG cantilever taken for a scan line perpendicular to the hinge. For each position of the laser, elastic signal is measured using a photodiode. Laser position is swept using a piezoelectric scanning stage. Fizeau interference fringes clearly appear and one can finely tune the position of the laser to maximize the reflected or Raman signal. \textbf{b}: Time modulation of the same reflected light intensity ($\lambda_{laser}$ = 532 nm) at a fixed position while a triangular quasi-static potential (dashed line) is applied between the gate and the suspended membrane. The gate-modulated optical signal shows quadratic festoon-like behavior. As suggested by the equation \ref{eq:forceV2}, optical signal is modulated at twice the actuation frequency. Depending on where the laser is focused, the festoons are pointing up (laser focused at a rising edge of a fringe - $\textcolor{vert}{\bullet}$) or pointing down (laser focused at a falling edge of a fringe - $\textcolor{blue}{\bullet}$). Such MLG cantilever offers the possibility to modulate either elastic or inelastic backscattered optical signals both spatially and electrostatically. }\label{fig6}
\end{center}
\end{figure}

 We now implement an interferometric detection of the motion, which would be maximum when the quantity $\chi_{opt}=\partial I_{reflected, Raman}/\partial h[\rho]$ is maximum, \textit{i.e.} at a fringe edge.
In our experiments, we manage to focus the laser beam at fringe edge, and far away for the hinge, in order to get large motion amplitude. 
Therefore, it is legitimate that $I_r$ is locally a linear function of $h(\rho, t)$ implying that for small deflection $\delta h(\rho,t)$, the variation of reflected light intensity (or Raman scattered light) under electrostatic actuation could be written as follows :
\begin{equation}
\Delta I_{reflected, Raman}[\rho,t] \sim \chi_{opt} \chi_{mec} B[h(\rho)_0] \ V(t)^2.
\label{eq:forceV2}
\end{equation}
The figure \ref{fig6}b shows the experimental values for reflected light intensity under triangular electrostatic actuation. 
When we apply a potential which is a linear function of the time, the associated change in reflected light intensity will be quadratic in time as suggested by the term in $V^2$ in the equation \ref{eq:forceV2}. 
Alternation of parabolic sections and peculiar sharp points is therefore due to the kinks of the $V(t)$ function (triangular signal).
We also observe that depending of the laser spot position, festoon-like behavior would be pointing up or down. 
Actually, this is due to the sign of $\chi_{opt}$ (cf. equation \ref{eq:forceV2}) which changes sign from one side of the Fizeau fringe to the other (rising or falling edge).
This effect allows us to calibrate precisely the magnitude of electrostatic actuation by analyzing the nonlinearities observed when more than one interfringe is swept (as shown is figure \ref{fig8}b and in reference\cite{Reserbat-Plantey2012}).
As we have shown previously, the typical order of magnitude for quasi-static actuation is about 1 nm.V$^{-2}$ for $\rho \sim$ 10 $\mu$m (thus, for 10V, motion amplitude at 10 $\mu$m from the hinge is about 100 nm).
 By varying the gate voltage applied on the MLG, its angle can be adjusted on a wide range (giving a local thickness variation far exceeding the wavelength of the incoming light).  its motion can be actuated and followed in real time from DC up to the tens of MHz range (mechanical resonance). Example of the optical detection of motion and stress of the membrane\cite{Reserbat-Plantey2012}.

To summarize, by combining the observed optical Fizeau interference to the possibility of adjusting the cavity length by electrostatic actuation of the MLG cantilever, it is possible to tune the optical resonance of the optical wedged cavity.
We now explore the potential of such interferometric enhancement with a graphene based hybrid system to measure Raman response of grafted molecules.

\section{Graphene as a tunable Raman platform for Molecular Detection}
\subsection{Adsorbed Species on Graphene : Hybrid Graphene systems}

One of the most promising aspects of graphene is the unique ability to combine its own properties with ones of a grafted specie.
Based on this feature, hybrid devices have been made to combine the electronic properties of graphene with the spin of a single molecular magnet\cite{Candini2011}.
Detection of single objects grafted on graphene still remains a crucial point to fully characterize hybrid devices and propose new ones.
Recently, the idea of using graphene as a substrate for Raman signal enhancement has been proposed \cite{Ling2010-2, Lopes2010}, revealing that graphene has distinctive properties for Raman signal enhancement of grafted chemical species.
The sensitivity goes up to very few amount of molecules ($\sim$ 10-100 for pyrene-substituted bis-phthalocyanine Terbium TbPc$_2$ case) \cite{Lopes2010,doi:10.1021/ja906165e} and is based on a chemical enhancement for due to charge transfer between the graphene and the molecules.
Graphene also has singular optical properties making it a semi-transparent membrane in the visible range \cite{Nair2008}. 
This property has been used for designing optical cavities made of graphene membrane lying on transparent SiO$_2$ layer, with silicon back mirror \cite{Blake2007}.
Optical interferences created by the cavity allows detection of graphene flakes when the cavity thickness is adjusted to be in resonant conditions (\textit{ie.} maximizing the optical contrast).
On that basis, interference enhancement of Raman scattering has been proposed \cite{Connell1980,Jung2010,Yoon2009} by tuning the SiO$_2$ thickness of samples made of supported or suspended graphene flakes parallel to the SiO$_2$/Si substrate.
That phenomenon of interference enhancement of Raman scattering (IERS) relies on the constructive interference between Raman scattered photons \cite{Connell1980,Fainstein1995}.

\subsection{Cavity enhancement of molecules adsorbed on graphene }
\begin{figure*}[htbp]
\begin{center}
		\includegraphics[width=\textwidth]{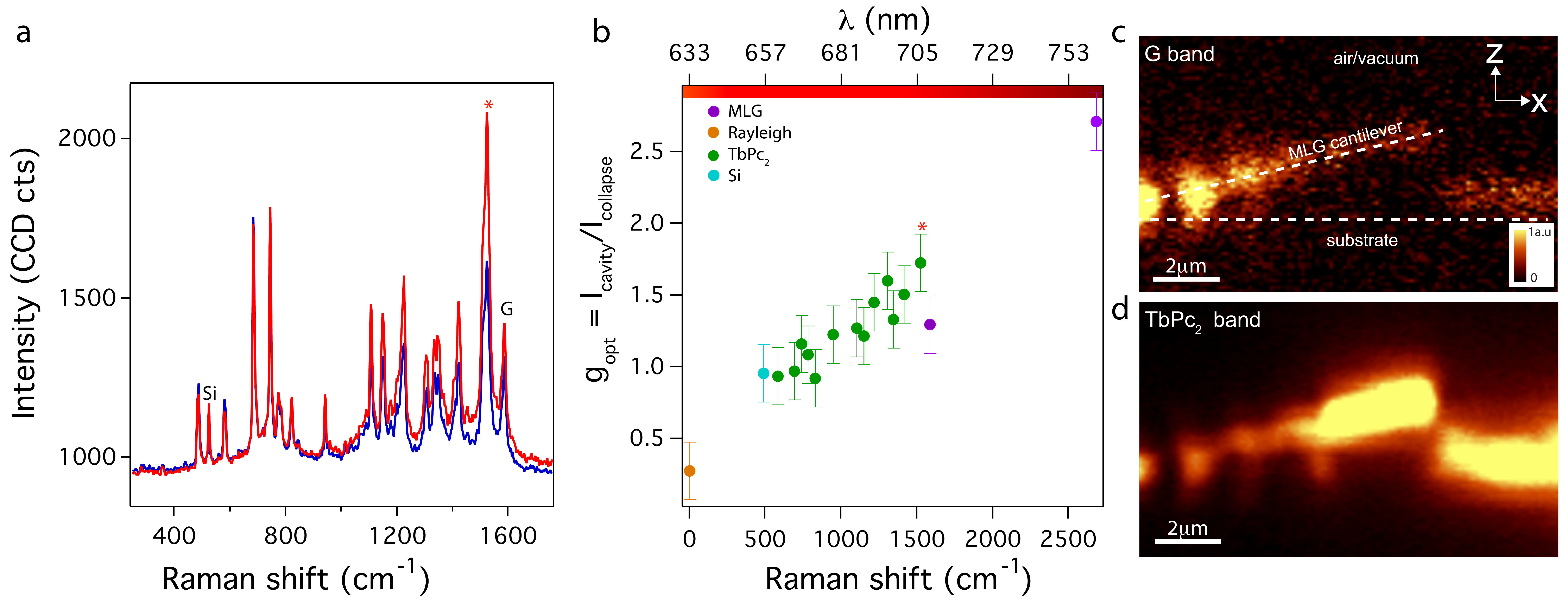}
	     \caption{\textbf{Modulation of Raman spectrum for generic adsorbed molecular species} \textbf{a}: Raman spectrum of hybrid system composed of TbPc$_2$ molecules grafted on the MLG cantilever top mirror. Laser spot is successively probing a bright fringe (red spectrum) for the MLG 2D band (red star) and onto the same system after mechanical collapse of the membrane on silicon (blue spectrum). \textbf{b}: optical gain $g_{opt} = I_{cavity}/I_{collapsed}$ as a function of the wavelength of Raman scattered photons. From a practical point of view, for each Raman mode $g_{opt}$ is the ratio of Raman intensity on top of Fizeau fringe with respect to Raman intensity on collapsed cantilever. Raman modes of the MLG cantilever (\textcolor{violet2}{$\bullet$}), substrate (\textcolor{lightblue}{$\bullet$}) and TbPc$_2$ (\textcolor{vert2}{$\bullet$}) are represented, as well as the Rayleigh peak (\textcolor{orange}{$\bullet$}). \textbf{c,d}: Raman mapping in depth scan configuration (x,z) of the G band Raman intensity (\textbf{c}) and 1515 cm$^{-1}$ TbPc$_2$ band (\textbf{d}) (indicated by \textcolor{red}{*} in \textbf{a} and \textbf{b}). Fizeau fringes are clearly observed for the modes.}\label{fig7}
\end{center}
\end{figure*}

Raman spectra of this hybrid system, as shown in figure \ref{fig7}a, indicates that the molecule is intact once deposited on the air wedge which is consistent with our previous work\cite{Lopes2010}.
Unlike the case where the graphene is parallel to the plane of the substrate\cite{Lopes2010}, our spatially resolved Raman mapping in the vertical plane (x,z) reveals Fizeau fringes for TbPc$_2$ Raman bands in addition to those Raman signal corresponding to the multilayer graphene membrane (see figure \ref{fig7}c-d).
By positioning the laser beam on the top of a bright interference Raman fringe of TbPc$_2$ (for example the Raman band at 1515 cm$^{-1}$), we measure an enhanced Raman signal of the molecule due to the interference effect. 
This enhancement depend on the finesse of the cavity, as suggested by the equation \ref{eq:Iramandetail} which is limited in our case by the absorbance of the two mirrors (silicon and graphene).
A higher cavity finesse would increase the gain of the Raman lines and also offer the possibility to completely extinguish a given Raman peak for instance. 
In order to quantify the optical gain in our cavity, we measure the Raman spectrum on the same system for two different configurations : (figure \ref{fig7}a) respectively on a bright Fizeau fringe (red curve) and where the membrane is collapsed on the substrate (blue curve).
In the second case, the optical thickness of the cavity is smaller than $\lambda/(2n_{SiO_2})$ for $\lambda \in[400; 800nm]$ ($h_{SiO2} \sim 180 nm $). 
The intensity ratio between the 2D peak on the top of a Fizeau fringe(identified by a red star in figure \ref{fig7}a )  and on the collapsed membrane is reported in figure \ref{fig7}b. 
The optical gain for each of Raman bands (molecule, silicon or membrane) follows an increasing dispersion behavior with the wavelength associated with each of the interference systems.
Since the gain differs even within the set of Raman peaks, it becomes possible to preferentially enhanced each Raman modes if one is able to change the thickness of the cavity by moving the laser or the membrane itself by electrostatic actuation, as we will
see in the next section.

\subsection{ AC Gate modulation of Raman lines of generic molecular species from molecular detection.}
\begin{figure}[htbp]
\begin{center}
		\includegraphics[width=0.48\textwidth]{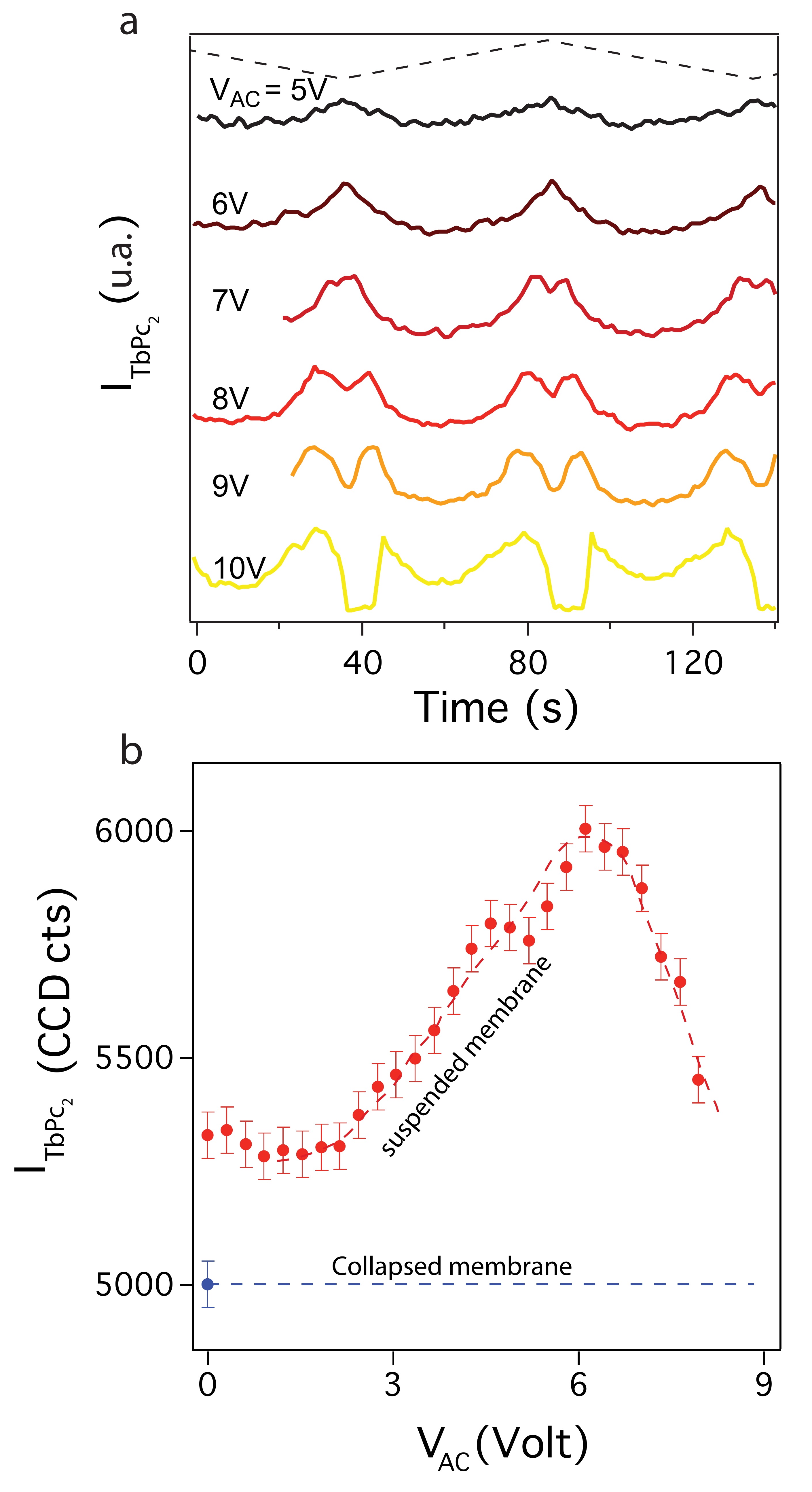}
	     \caption{\textbf{Time modulation and cavity enhancement of Raman lines of generic molecular species adsorbed on the graphene MLG. } \textbf{a}: Time trace of the 1515 cm$^{-1}$ TbPc$_2$ Raman band while MLG cantilever is electrostatically actuated with an external triangular potential $\delta V_{AC}$ (dark dashed line). $I_{TbPc_2} (t)$ shows festoon-like behavior which is distorted as $\delta V_{AC}$ becomes larger that 7V. \textbf{b}: Evolution of the grafted molecule same Raman signal $I_{TbPc_2}$ as a function of the static actuation. Red dots are recorded for suspended movable MLG cantilever while blue dots refer to the same position after MLG mirror has been collapsed. The cavity is in resonant conditions for $\delta V_{AC} \sim$ 6V for that particular Raman mode at 1515 cm$^{-1}$.}\label{fig8}
\end{center}
\end{figure}
As described previously, the Raman response TbPc$_2$ molecules grafted on the suspended MLG membrane is affected by the interference phenomenon. 
By applying the AC voltage, the thickness of the cavity - and hence the position of TbPc$_2$ fringes - is periodically shifter (cf. equation \ref{eq:forceV2}).
Ajusting the cavity length offers two features that are unique for generic Raman studies and potentially very useful to detect tiny Raman signal emitted by small amount of molecules : 
firstly the possibility to adjust the molecule under the Laser probe up to a resonating position for maximizing the Raman signal of the molecule.
 Secondly, the possibility to modulate in time Raman intensities, in a synchronized way with the AC voltage, a feature which potentially allows lock-in detection at a specific wavelength.
 To validate these two features, we have measure the response of the Raman spectra with respective application of AC and DC voltage (fig. 8 a and b, respectively).  
 
For an applied AC sawtooth signal (\ref{fig8}a.), we systematically detect a festoons-like responseof the reflected light on the membrane which is similar to what was observed for Graphene Raman signal, as was shown in figure \ref{fig8}a.
When $V_{AC}$ is gradually increased above  $V_{AC}=6V$ , the thickness of the cavity thickness span than  $\delta h=\lambda/2n_{0}$, which means that one full Fizeau interference fringe has been passed through which explain the dip which appealer
In figure \ref{fig8}b, Raman intensity of TbPc$_2$ mode (1515 cm$^{-1}$) is plotted versus the applied voltage and the measured Raman signal compared to a collapsed immobile membrane. 
The optical gain is maximum when the thickness the cavity was adjusted to have a bright fringe (constructive interference) of the molecule Raman mode (for $V_{AC}>$7V).
The optical cavity is thus modulated electrostatically to adjust its thickness, which maximize the Raman line intensity. 
This important result allows synchronized detection methods where the excitation signal is locked with the recorded frequency of Raman spectrum, for example. 
Such a technique would lower the detection threshold of adsorbed molecules on the membrane, tending towards the detection of a single object. 
Thanks to the electrostatic modulation of the cavity, ones can select the Raman modes enhanced by a constructive interference. 
By using this phenomena, we can thus inverse the intensity ratio between the selected modes and all others Raman signals. 
For example, we can selectively reject the fluorescence signal thus improving the signal to noise ratio of Raman spectra. 

\section{Conclusion}

We have characterized the properties of simply clamped graphene cantilevers that acts as a dual mechanical and optical resonators: both NEMS and Fabry-Perot like cavity. 
This combination realizes an integration of a Raman based platform for the detection of molecular grafted species. 
When overhanging oxidized silicon substrates, graphene cantilevers implement an optical cavity with finesse in the range 5-10. 
Thickness variation within the air wedge between the graphene membrane and silica induces equal spacing fringes (Fizeau fringes) in the reflected and scattered light. 
We have shown that such optical cavity exhibits optical interferences pattern (Fizeau fringes) visible for both incoming laser beam and Raman scattered photons.
The fact that both elastic and inelastic scattering response give rise to a series of interference patterns which specific signature and interfringes , allows splitting each Raman modes coming from different chemical origin and vertical position  
Using that feature, we have been able to selectively reject the graphene Raman signals from the silicon one for example.
Such tuning is possible either by moving the laser spot along the Fizeau fringes pattern, or by electrostatically actuating the cavity.
Moreover, it constitutes a case-study system to get a comprehensive picture of optical interferences within graphene-based optical cavities.
When graphene surface is  functionalized with $\pi$-stacked molecules, the Raman signal intensity is modulated by actuation of the cavity length with gate voltage, and shows an enhancement which is maximum when the characteristic wavelength of the associated Raman band is at resonant conditions. 
The achieved time modulation of Raman paves the way for lock-in type detection for single molecule Raman spectroscopy.
The extension of that fine space and time tuning to any arbitrary Raman signal originating from chemicals grafted on the graphene show that our original platform offers a promising device for highly sensitive Raman experiments on grafted chemicals on graphene. Further works are underway to take profit of this system for the studies by Raman spectroscopy of single molecular systems.

\acknowledgments

We thank D. Basko, E. Bonet, A. Candini, L. Del-Rey, E. Eyraud, J. Jarreau, M. Lopes, C. Schwarz, M. Urdampilleta and W. Wernsdorfer for discussions and help in realizing the experimental setup. We also thank Nanochemistry and Nanofab facilitie at Néel Institute. This work was supported by the Agence Nationale de la Recherche  (ANR projects : MolNanoSpin, Supergraph, Allucinan and Trico), European Research Council (ERC advanced grant no. 226558), the Nanosciences Foundation of Grenoble and Region Rhône-Alpes. M.R. thanks the priority program PP 1459 of the German Science Foundation (DFG) for generous support.  

\bibliography{BIB-cavity}

\section{Supplementary Material}
\subsection{Methods} 
Multilayered graphene flakes are deposited on 280 nm thick oxidized silicon wafer by micro-mechanical exfoliation of Kish graphite. 
Electrical contacts are made using deep UV lithography and e-beam deposition of 50 nm Au electrodes. Samples are suspended by etching (in buffered fluorhydric acid solution 1:3 HF/NH$_{4}$F))  transferred in isopropyl alcohol and dried using CO$_{2}$ critical point drying. 
Reflectivity and Micro-Raman spectroscopy were performed with a commercial Witec Alpha 500 spectrometer set with a dual axis X-Y piezo stage in a back-scattering/reflection configuration allowing a spatial imaging of our samples.
Raman spectra are recorded on each pixel and the integrated intensity of a chosen mode is displayed on a false color scale (see for instance figures \ref{fig2}b or \ref{fig1}c). 
This microRaman is mounted on a  confocal microscope equiped with a Nikon 100x objective with a numerical aperture of 0.95, conferring a spot size $\le$ 500 nm at a focus depth of 1 $\mu$m.
Two laser excitation wavelengths are used, 633 nm (He-Ne) and 532 nm (Solid state Argon diode).
All Raman spectra have been recorded by a CCD camera with 1 kHz bandwidth.
The power of the laser beam is kept below 300 $\mu$W to prevent any damage on the MLG film. 
Non-contact AFM is performed on a collapsed region of the same flake to obtain the MLG thickness. 
Additional measurements using optical confocal microscopy give an estimate of the flake geometry and of the angle of the air wedge, which is further confirmed using SEM microscopy with grazing incidence. 
Ellipsometry simulation is performed using software from NanoCalc thinfilm reflectometry system by Ocean Optics.

Multilayered Graphene flakes (MLGs) are deposited on oxidized silicon wafer by micro-mechanical exfoliation\cite{{Novoselov2004},{Novoselov2005}} of Kish graphite on a 280 nm SiO$_2$-Silicon wafer. 
After exfoliation, pattern of electrically connected leads made of 50nm thick gold is deposited on top of the flakes by using deep-UV lithography followed by electron beam evaporation and lift-off. 
Silicon oxide supporting the graphene flake is etched on approximately 90 nm by dipping the sample in a solution of buffered hydrogen fluoride (concentration 1:3 HF/NH$_4$F) during 65 seconds and drying the sample using CO$_2$ critical point drying.
MLG films (typically 10-100 MLs) exhibit sufficient rigidity to stay free standing without collapsing along micron size distances even when they are simply clamped by one side. 
We identify the flakes interesting for our study by selecting those which have one side clamped by a gold lead while the other extremity is free standing and tend to stick-up after the fabrication process (see Fig.\ref{fig:fig1}c) thus realizing air wedge structure with air thickness ranging between 0.1 and 2 microns. 
A clear iridescent pattern of fringes can be seen on a small fraction (ca. 5\%) of these partly suspended flakes which tells that a specific range of flake thicknesses is required for observing this effect. 
A typical density of 10 flakes per cm$^2$ show this effect. 
Experiment has been performed on approximately 100 samples.

The molecules deposited on graphene are pyrenyl-substituted heteroleptical bis(phthalocyaninato) terbium(III) complexes, referred to as TbPc2 in this publication\cite{doi:10.1021/ja906165e}.
TbPc$_2$ molecule is deposited onto the sample at the very end of the fabrication process. 
We use spin drop casting, at high rotational speed, with no other process.
Almost 90 $\%$ of the suspended cantilevers remain suspended after that step.
More details of the deposition method are in reference\cite{Lopes2010}.

\subsection{Interference pattern along the air wedge}

A simplified simulation is presented in figure \ref{fig3}.
It considers only a two wave interference, involving the reflections on the MLG and Si planes respectively. The protocol consists in varying the focused spot position of an incoming Gaussian beam, and measuring the overlap of the reflected field with the spatial profile of the incoming field, that also matches the detection mode. 
The reflected field is estimated from the complex sum of the fields reflected by the two planes, while taking into account the field reflection and transmission coefficients on each mirror.
  Note that no spatially incoherent scattering has been considered in this very simple approach, which explains the absence of the bright features observed experimentally at the plane location.
   However, the general characteristics of the Fizeau fringes can be reproduced (see Fig. 1f). Note also that in order to obtain a correct fringe pattern, one has to take into account the phase curvature of the fields, as a simple intensity calculation is not sufficient for reproducing the observed patterns. 

\begin{figure}[htbp]
\begin{center}
		\includegraphics[width=0.48\textwidth]{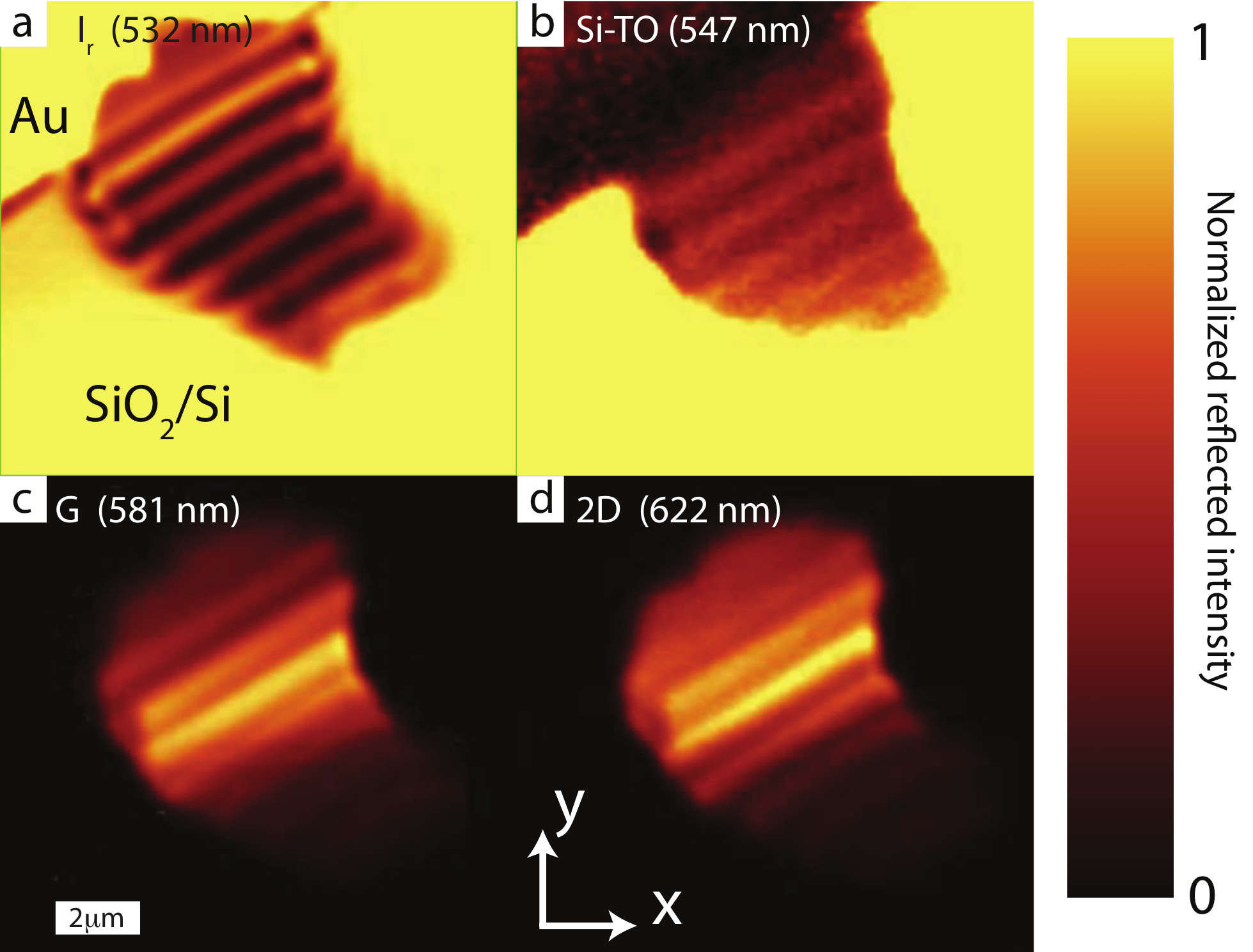}
	     \caption{\textbf{Raman Fizeau fringes along MLG optical cavity.} Light intensity mapping for \textbf{a} reflected light, \textbf{b} Raman Si-TO band, \textbf{c} Raman G and \textbf{d} 2D bands of MLG membrane. For each Raman line, the corresponding photon wavelength is indicated. Due to the wedge geometry, confocal mapping at constant height does not allow to be focused over the whole membrane, as shown in the global intensity variations in \textbf{c-d}. }\label{figSI1}
\end{center}
\end{figure}

\begin{figure}[htbp]
\begin{center}
		\includegraphics[width=0.48\textwidth]{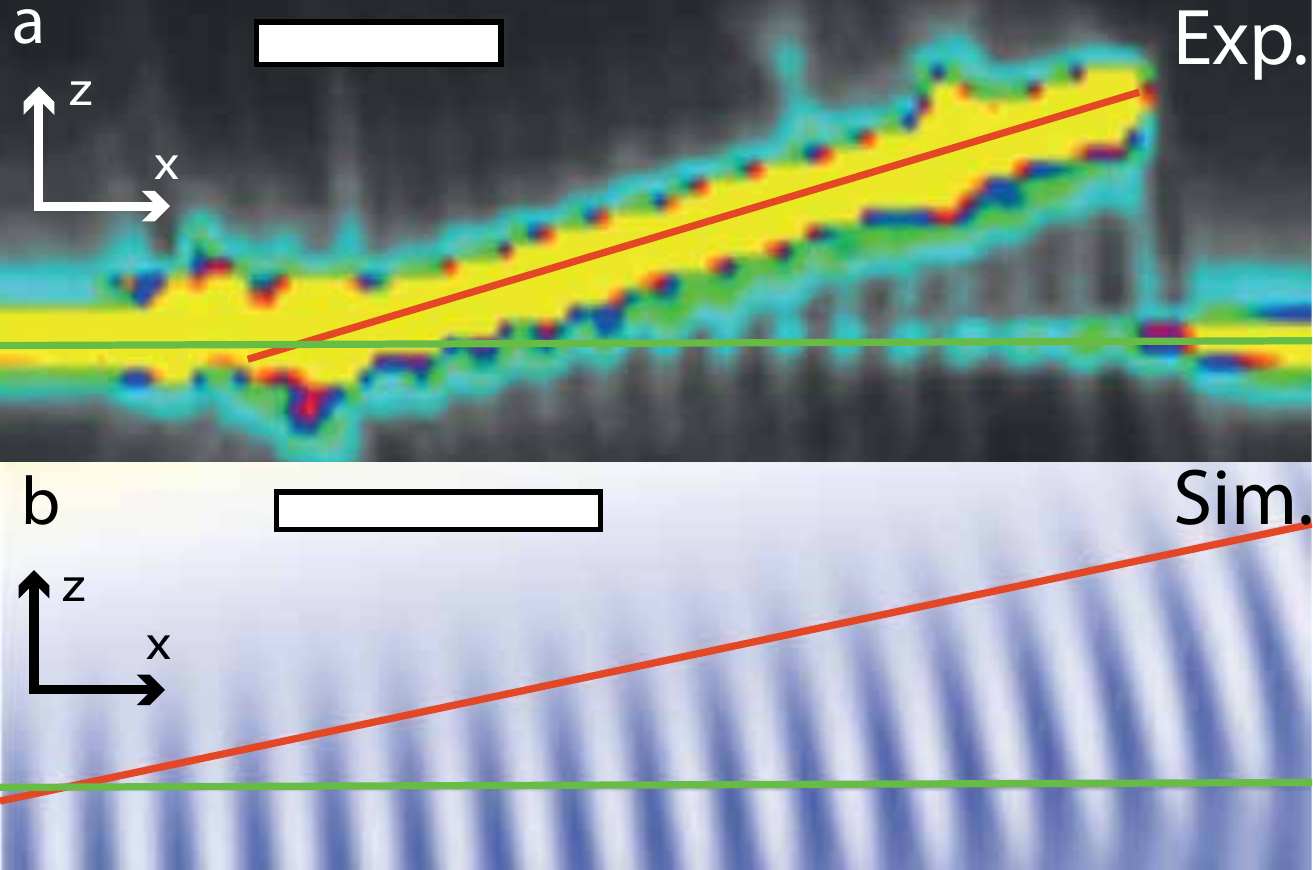}
	     \caption{ \textbf{Interference profile of a MLG air wedge cavity : comparison between experiments and simulations.} \textbf{a} Spatially resolved reflectance map in the cross sectional (X,Z) plane at 532 nm for a typical sample. Reflected laser light was recorded using scanning confocal microscopy at different focus height. \textbf{b} Interference profile of a MLG wedged cavity obtained by simulation using coherent superposition of two waves. Reflected intensity is reported in color scale, arbitrary units, white being highest intensity. This has been calculated for a wavelength of 500 nm, a numerical aperture of 0.9, reflection and transmission coefficients of 50 percent in intensity and an angle of $\alpha=12^{\circ}$. Red line corresponds to the MLG cantilever position and  green line to the interface between silicon oxide and vacuum. Scale bars are 5 $\mu$m.}\label{figSI2}
\end{center}
\end{figure}

\end{document}